\documentclass[prb,aps,epsfig,twocolumn,showpacs]{revtex4-1}
\usepackage{color}
\usepackage{epsfig,amssymb,amsmath,latexsym}
\usepackage{float}
\usepackage{color}
\usepackage[colorlinks=true,linktoc=page,linkcolor=magenta,citecolor=magenta]{hyperref}

\newcommand\bea{\begin{eqnarray}}
\newcommand\eea{\end{eqnarray}}
\newcommand\beq{\begin{equation}}  
\newcommand\eeq{\end{equation}}

\newcommand\ie{{\it{i.e.}}}

\begin{document}
\textheight=23.8cm

\title{Conductance, Valley and Spin polarization and Tunnelling magneto-resistance in ferromagnetic$-$normal$-$ferromagnetic junctions of silicene}
\author{Ruchi Saxena$^{1}$, Arijit Saha$^{2}$ and Sumathi Rao$^{1}$}
\affiliation{\mbox{$^1$} {Harish-Chandra Research Institute, Chhatnag Road, Jhunsi, Allahabad, Uttar Pradesh,  211 019, India} \\
\mbox{$^2$}{Institute of Physics, Sachivalaya Marg, Bhubaneswar, Orissa, 751005, India}
}

\date{\today}

\begin{abstract}
We investigate charge conductance and spin and valley polarization along with the tunnelling magneto-resistance (TMR) in silicene junctions 
composed of normal silicene and ferromagnetic silicene. We show distinct features of the conductances for parallel and anti-parallel spin
configurations and the TMR, as the ferromagnetic$-$normal$-$ferromagnetic (FNF) junction is tuned by an external
electric field. We analyse the behavior of the charge conductance and valley and spin polarizations in terms of the independent 
conductances of the different spins at the two valleys and the band structure of ferromagnetic silicene and show how the conductances are 
affected by the vanishing of the propagating states at one or the other valley. In particular, unlike in graphene, the 
band structure at the two valleys are independently affected by the spin in the ferromagnetic regions and
lead to non-zero, and in certain parameter regimes, pure valley and spin polarizations, which can be tuned by the external electric field.
We also investigate the oscillatory behavior of the TMR with respect to the strength of the barrier potential (both spin-independent and
spin-dependent barriers) in the normal silicene region and note that in some parameter regimes, the TMR can even go from  positive to 
negative values, as a function of the external electric field.
\end{abstract}

\maketitle

\section{Introduction}

A close cousin of graphene, silicene has been attracting a lot of attention in recent times both theoretically
and experimentally~\cite{MEzawa4,siliceneexp1,siliceneexp2,siliceneexp3}, due to the possibility of new applications, 
given its compatibility with silicon based electronics. Although earlier theoretical analyses~\cite{CCLiu1,MEzawa2,MEzawa3}
had predicted the possibility of silicene and even germanene, stanene, etc, interest in this subject rose after experimentalists 
observed hexagonal structure in silicene sheets deposited on a silver substrate~\cite{siliceneexp1,siliceneexp2,siliceneexp3}. 
Unlike graphene, silicene does not have a planar structure; instead it forms a buckled structure due to the large atomic radius of silicon,
resulting in a band gap at the Dirac point~\cite{CCLiu1}. Further, it turns out that such a band gap is tunable by an external 
electric field applied perpendicular to the silicene sheet~\cite{Drummond,MEzawa3}. 
Thus, from the point of view of applications, one of the drawbacks of graphene in making transistors is overcome by silicene, 
and very recently, a silicene based transistor has been experimentally realised~\cite{LTao}. 
Therefore, with the possibility of silicene based electronics, experimental interest in silicene remains quite high.

Theoretical interest in silicene soared when it was realised that one could have topologically non-trivial phases in silicene,
tuned by the external electric field~\cite{Drummond,MEzawa3}. Graphene and silicene have similar band structures and the 
low energy spectrum of both are described by the relativistic Dirac equation~\ie, both have the Dirac cone band structure 
around the two valleys represented by the momenta ${\bf K}$ and ${\bf K\rq{}}$. However, in silicene, 
the spin-orbit coupling is much larger than in graphene~\cite{GianG,CCLiu1,MEzawa3}. This is an important difference which  
causes the Dirac fermions to become massive.
Furthermore, because of the buckled structure in silicene, the two sub-lattices respond differently to an externally applied
electric field. This means that the Dirac mass term becomes tunable~\cite{MEzawa3},  
and hence allows for the mass gap to be closed at some critical value of the electric field and then reopened. 
The phases on the two sides of the critical value where the gap is closed are different, with one of them
being topologically trivial and the other being topologically non-trivial~\cite{MEzawa2,MEzawa3,MEzawa5}. 
Hence, under suitable circumstances, silicene can be a quantum spin Hall insulator with topologically protected 
edge states~\cite{CCLiu2,MEzawa1}. 

In recent years, spin based electronics or spintronics has become a prominent field of research both theoretically and experimentally
~\cite{RevModPhys.76.323}. The upsurge of activity in this area is essentially due to the realisation that devices based on the spin degree 
of freedom are almost dissipationless unlike those based on the charge degree of freedom. The possibility of using silicene as a spintronic 
device has been reported very recently in Ref.~[\onlinecite{WangRuge,MEzawa6,XingTaoAn,ABansil,fmpeeters2}] due to its strong spin-orbit coupling. 
Another important quantity in spintronics is the tunneling magneto$-$resistance (TMR) which occurs at junctions between materials~\ie, 
ferromagnet$-$normal metal$-$ferromagnet (FNF) junctions. The resistance of the junction is different for parallel and anti-parallel spin 
configurations, and this difference in resistance can be experimentally measured.

In this paper, we study charge conductance and spin and valley polarizations along with the TMR in silicene junctions, 
in particular the  FNF junctions, as the external electric field  is tuned through the system. We model our FNF setup within 
the scattering matrix formalism~\cite{SDatta}. A similar setup  has  been utilised in graphene~\cite{JZou} to study TMR, while 
valley polarization has been studied in a normal$-$ferromagnet$-$normal (NFN) junction 
in silicene~\cite{TYokoyama}. In Ref.~{\onlinecite{TYokoyama}}, the author has investigated the conductances and valley polarization of a 
NFN junctions in silicene.  Spin and valley textures of the particle-hole excitations  due to the addition of external fields is another important
issue in silicene and recently, analytical and numerical results for the dispersion of the plasmons has been studied~\cite{fmpeeters}.
However, conductances and TMR based on silicene FNF junction have not yet been considered in the literature.

The remainder of this paper is organized as follows. In Sec.~{\ref{sec:II}}, we present our model and band structure for the ferromagnetic 
silicene (FS). In Sec.~{\ref{sec:III}}, we describe the scattering matrix for the FNF junction to compute the two terminal charge conductance ($G_c$), 
valley and spin polarizations ($\mathcal{P}_v, \mathcal{P}_S$) and the TMR. In Sec.~{\ref{sec:IV}}, we present our numerical results for the 
conductances, spin and valley polarizations and TMR in the FNF set up for various parameter regimes. Finally in Sec.~{\ref{sec:V}}, we present 
the summary of our numerical results followed by the conclusions.

\section{Model and band structure} {\label{sec:II}}

%
\begin{figure}
\centering
\includegraphics[width=1.0\linewidth]{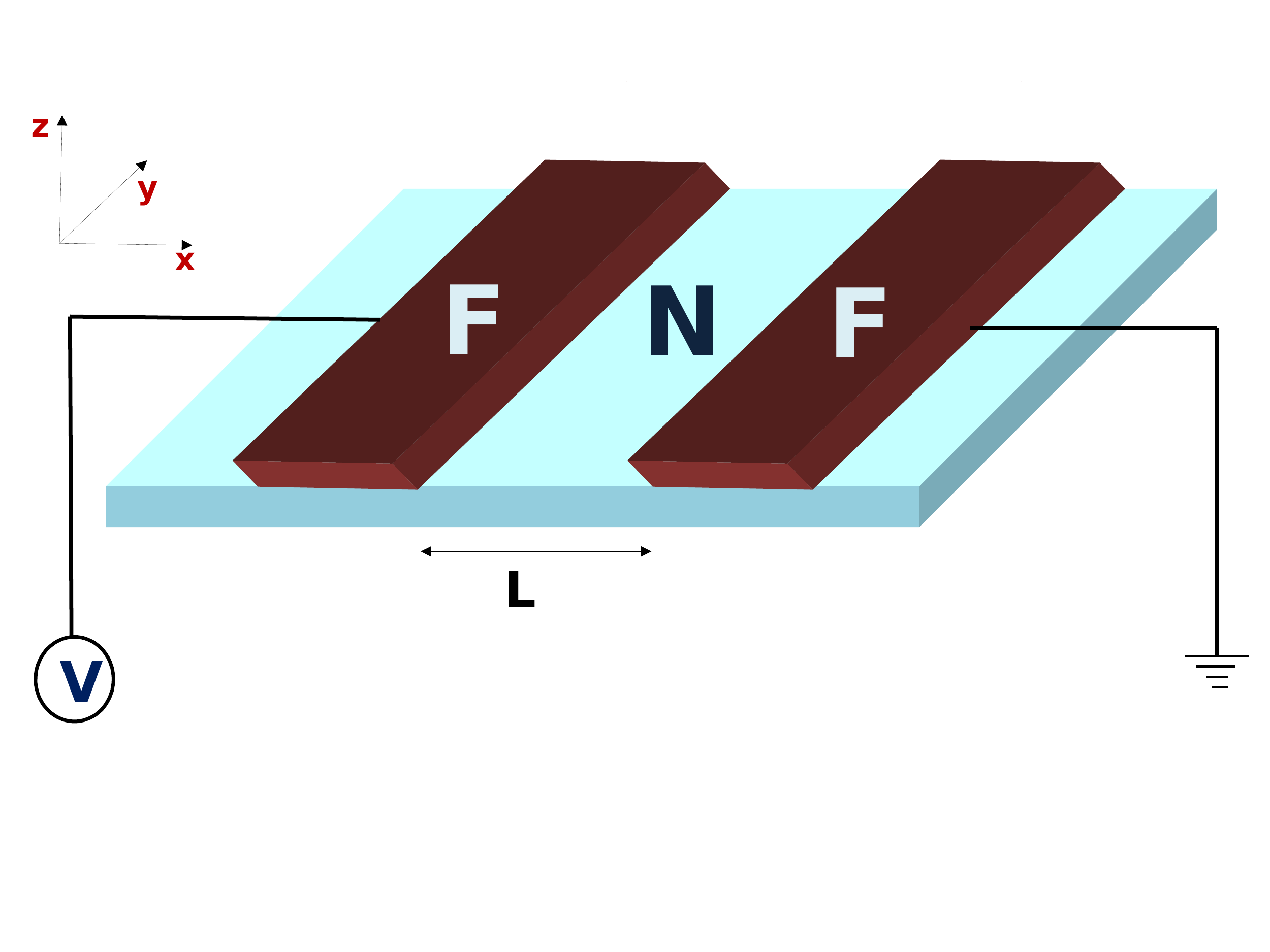}
\caption{(Color online) Schematic of the  FNF junction in silicene in which two ferromagnetic patches (dark brown, dark grey) have been 
deposited on two sides of a normal silicene sheet (cyan, light grey) to induce ferromagnetism in it. Here $ L$ is the 
length of the normal silicene region.}
\label{Fig1}
\end{figure}
%

We study the FNF junction set up in silicene as shown in Fig.~\ref{Fig1}. Ferromagnetism is induced in silicene by the proximity effect 
when it is placed in proximity with a magnetic insulator, which we model by the following Hamiltonian given by~\cite{TYokoyama}
\beq
H = \hbar v_F (\eta k_x \hat{\tau}_x - k_y \hat{\tau}_y) + (e l E_z - \eta \sigma \lambda_{SO})\hat{\tau}_z + V(x)- h(x),
\label{hamiltonian}
\eeq
where $v_F$ is the Fermi velocity of the charge carriers in silicene, $e$is the charge of the electron and $\eta, \sigma$ correspond to the valley and spin indices and
$ \hat{\tau}$ corresponds to the sublattice (pseudospin) Pauli matrices. $\lambda_{SO}$ is the parameter that specifies the spin-orbit coupling 
in silicene. Due to the buckled structure of silicene, the atoms in the two sublattices respond differently to an externally applied electric field 
$E_z$~\cite{MEzawa3}. Thus $l E_z$ is the potential difference between the two sublattices A and B due to this applied electric field 
where $l$ is the separation between the two sublattices. Hence, the potential difference is a tunable parameter and can be tuned by an 
external electric field~\cite{MEzawa3}. Also, when the Fermi energy is close to the Dirac point, 
at  the critical electric field $E_{z}^{c}=\eta \sigma \lambda_{SO}$, one of the valleys in silicene is 
up-spin polarized and the other one is down-spin polarised~\cite{MEzawa3}. 
Here $\eta=\pm 1$ denotes the ${\bf K}$ and ${\bf K}^{\prime}$ valleys respectively and $\sigma=\pm 1$ denotes 
the spin indices. $V(x)$ denotes the profile for the potential barrier in the normal silicene region and $h(x)$ corresponds to the 
exchange splitting, or the energy difference between the up and down spin electrons in the FS regions. Note however, that 
in real materials the proximity of a ferromagnet to silicence can actually change the band structure of silicene itself.
In that case, the only way to proceed will be to perform first principles calculations as have been done in graphene~\cite{hxyang}.

%
\begin{figure}[h]
\centering
\includegraphics[width=1.1\linewidth]{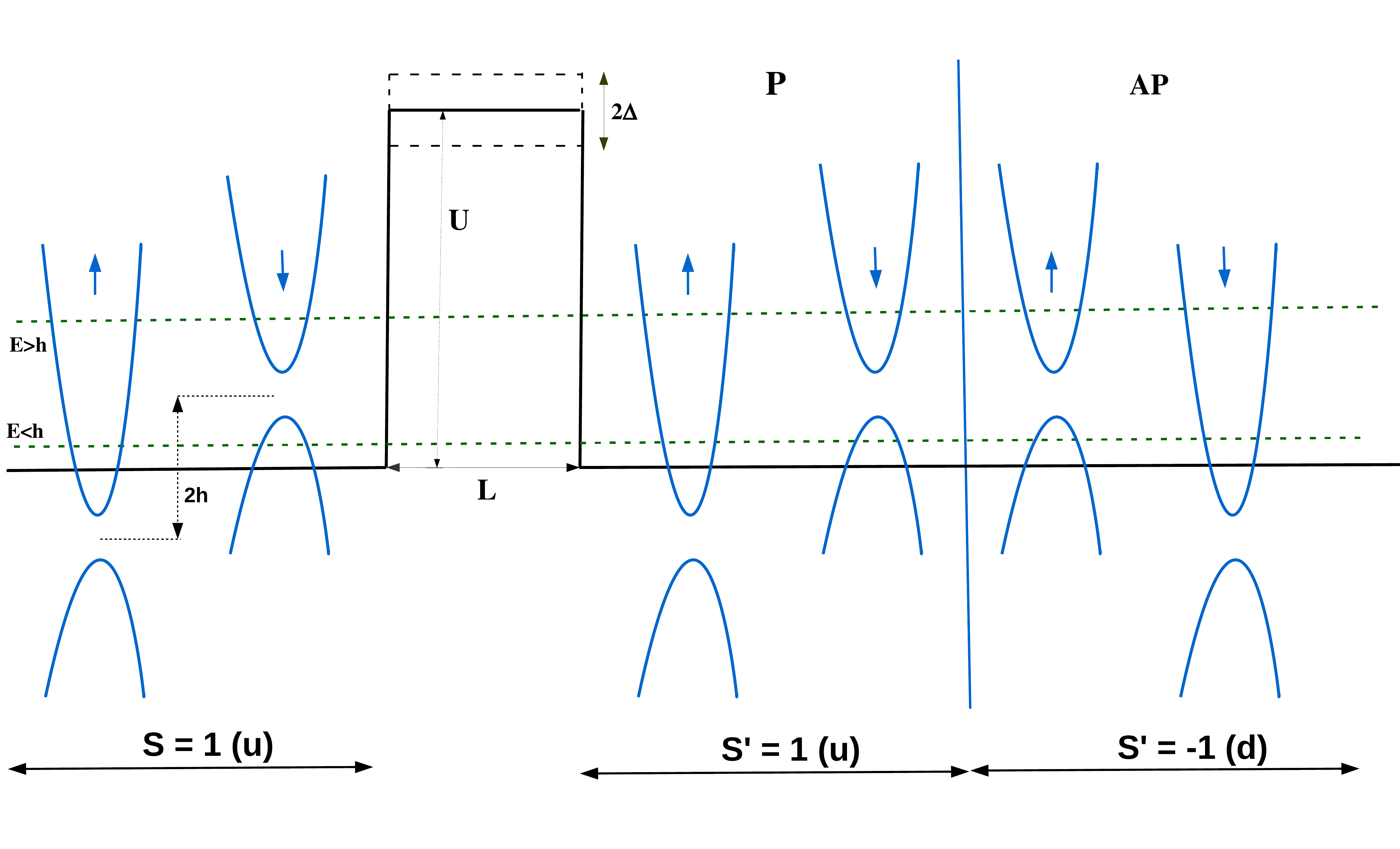}
\caption{(Color online) Schematic of the configurations with P ($uu$) and AP ($ud$) spin polarizations for one of the valleys of the 
FNF silicene junction.}
\label{Fig2}
\end{figure}
%

We now consider the geometry shown in Fig.~\ref{Fig1} and assume that the system is translationally invariant along the $y$ direction.
The interface between the normal and the FS are located at $x = 0$ and $x = L$ where $L$ the length of the
normal silicene region sandwiched between the ferromagnetic patches. 
Here $V(x) = U \Theta(x) \Theta(L-x)$ is the profile of the potential barrier modelled in the normal silicene region and 
$h(x) = h [\sigma s \Theta(-x) + \sigma' s' \Theta(x-L)]$ denotes the exchange field or Zeeman field in the two ferromagnetic regions 
with  $s=s\rq{}$ corresponding to the parallel (P) configuration and $s=-s\rq{}$ corresponding to the 
antiparallel (AP) spin configurations of magnetization respectively. We show the schematic of 
up and down spin in left region (for $s=1$) and in right region for parallel ($s'=1$ or P or $uu$) and antiparallel ($s'=-1$ or AP or $ud$) configurations 
in FNF junction for both $E<h$ and $E>h$ regions in Fig.~\ref{Fig2}. The orientation of the exchange field in the left region
is kept fixed by keeping $s=1$ and then $s'=1$ and $s'=-1$ in the right region corresponds to the parallel (P or $uu$) and antiparallel (AP or $ud$) 
configuration respectively. Note that the $E<h$ line crosses the same band in the third region for the P configuration, but crosses the other band 
for the AP configuration. As reported earlier~\cite{JZou}, this feature gives rise to negative TMR in graphene. In silicene, also, the same 
feature is responsible for negative TMR which we discuss later in Sec.~{\ref{sec:IV}}.

%
\begin{figure}[h]
\centering
\includegraphics[width=1.0\linewidth]{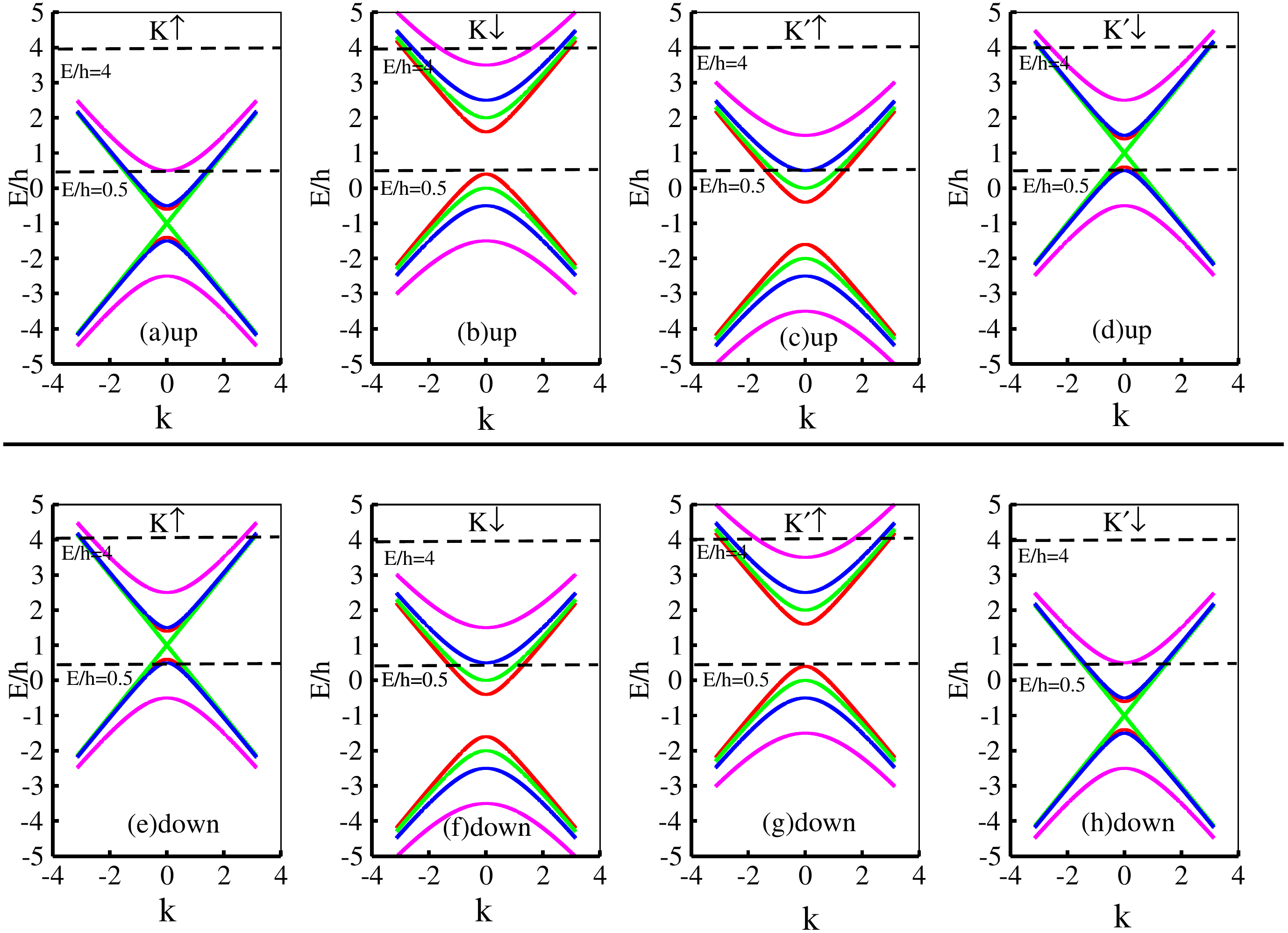}
\caption{(Color online) (a) to (d) gives the schematic of the band structure for $s=+1$, at $\bf K$ (for both $\uparrow$ and $\downarrow$ spin) and 
${\bf K}^{\prime}$ (for both $\uparrow$ and $\downarrow$ spin) valleys for ferromagnetic silicene for four different values of the 
dimensionless parameter $E_z/h$ (red (0.1), green (0.5), blue (1.0) and magneta (2.0)). On the other hand (e) to (h) gives the same for $s=-1$.
We use these diagrams to  qualitatively explain the dependence of conductances on the electric field as mentioned in the text.
}
\label{Fig3}
\end{figure}
%

In Fig.~\ref{Fig3} we demonstrate the spin polarization of both the $\bf {K}$ and $\bf {K^{\prime}}$ valleys  for 
FS with the magnetisation directions up ($u$ defined by $s=+1$) and down ($d$ defined by $s=-1$) via an energy band diagram.
The  diagram clearly shows  that different spin orientations behave differently in the 
$\bf {K}$ and $\bf {K^{\prime}}$ valleys at different values of the exchange field $h$. To visualize this picture, we fix 
$\lambda_{SO}/h=0.5$ and show the dispersion for $\bf {K \uparrow}$, $\bf {K \downarrow}$, $\bf {K^{\prime}\uparrow}$ 
and $\bf {K^{\prime} \downarrow}$ for four  different values of the tunable parameter $E_{z}/h$, for the cases $u$ and $d$.
Note that we use $\uparrow$, $\downarrow$ to denote the spins of the incoming (also reflected and transmitted) charge carriers
and we use $u,d$ to denote the orientation of magnetic exchange. For the $u$ or $s=+1$ case, 
at $E_{z}/h=\lambda_{SO}/h=0.5$
both $\bf {K}\uparrow$ and $\bf {K^{\prime}}\downarrow$ have a vanishing band gap. On the other hand, for $E_{z}/h=0.1$,  $E_{z}/h=1$, and
$E_z/h=2$, the valleys at $\bf {K}\uparrow$, $\bf {K}\downarrow$, $\bf {K^{\prime}\uparrow}$, $\bf {K^{\prime}}\downarrow$ are all gapped.
Also note that due to the exchange splitting $h$,
the $\bf {K^{\prime}}$ valley is shifted upwards for the 
$\downarrow$ spin while the $\bf {K}$ valley is shifted downwards for the $\uparrow$ spin. 
The case for $d$ or $s=-1$ is the other way around. Hence, it is clear that unlike in graphene, 
the contributions to the conductances from various spin configurations will not be identical for the ${\bf K}$ and ${\bf K}^\prime$ valleys.



\section{The scattering matrix}  {\label{sec:III}}

We model our FNF setup within the scattering matrix formalism~\cite{SDatta} where we match the wave functions at each ferromagnet$-$normal 
interface to obtain the scattering matrix and find the conductances and the TMR .
The wave functions for the valley $\eta$ in each of the three regions, $x<0$, $0<x<L$ and $x>L$ can be written as 
\bea
\psi_i &=& a_i \frac{\mathrm{e}^{i k_{ix}x}}{\sqrt{2 E \tau_i}}\left(\begin{array}{cccc}\eta k_i S_i \mathrm{e}^{i \eta \theta_i} \\ 
\tau_i \end{array} \right)  \nonumber \\    &+& b_i\frac{ \mathrm{e}^{-i k_{ix}x}}{\sqrt{2 E \tau_i}}
\left(\begin{array}{cccc}-\eta k_i S_i\mathrm{e}^{-i \eta \theta_i} \\ \tau_i \end{array}
\right)\ ,
\label{WF1}
\eea
where $a_1=1, b_1=r$  for $x<0$, $a_2=a, b_2=b$ for $0<x<L$ and $a_3=t,b_3=0$ for $x>L$.   
Note that we keep track of the sign of the charge carriers by including the index $S_i$ in all the regions (the sign of the charge carriers changes 
from electron-type to hole-type when $S_i$ is negative in any region). This actually happens 
for the anti-parallel configuration when energy of the incident charge carrier is below the induced magnetic field energy \ie~($E<h$).  
This charge reversal actually qualitatively changes the conductances, as was shown in graphene~\cite{JZou}.

We obtain the scattering matrix both for $E>h$ and $E<h$ by matching the wave functions (see Eq.(\ref{WF1})) at $x = 0$ and $x = L$ 
and solving Eq.(\ref{matrix}) numerically.
\begin{widetext}
\bea  \begin{bmatrix} \frac{-\eta k_1 S_1\mathrm{e}^{-i \eta \theta_1}}
{\sqrt{2 E \tau_1}} & -2 \eta k_2 S_2 \mathrm{e}^{i \eta \theta_2} & \eta k_2 S_2 
\mathrm{e}^{-i \eta \theta_2} & 0 \\ \sqrt{\frac{\tau_1}{2 E }} &  -\tau_2 & 
-\tau_2  &  0 \\ 0 & 2 \eta k_2 S_2 \mathrm{e}^{i \eta \theta_2} \mathrm{e}^{i k_{2 x} L}
 & -2 \eta k_2 S_2 \mathrm{e}^{-i \eta \theta_2} \mathrm{e}^{-i k_{2 x} L} 
& \frac{\eta k_3 S_3\mathrm{e}^{i \eta \theta_3} \mathrm{e}^{-i k_{3 x} L}}{\sqrt{2 E \tau_3}}\\ 0 & \tau_2 \mathrm{e}^{i k_{2 x} L}
 & \tau_2 \mathrm{e}^{-i k_{2 x} L} & -\mathrm{e}^{i k_{3 x} L}
\sqrt{\frac{\tau_3 }{2 E}}\end{bmatrix}  \left[ \begin{array}{c} r \\  a\\ b \\ t  \end{array} \right] = \left[ \begin{array}{c} \frac{-\eta k_1 S_1\mathrm{e}^{i \eta \theta_1}}
{\sqrt{2 E \tau_1}} \\  -\sqrt{\frac{\tau_1 }{2 E}}\\ 0\\0 \end{array} \right] 
\label{matrix}
\eea
\end{widetext}
Further, 
\bea
 k_i &=& \sqrt{{E_i}^2 - {(e l E_z - \eta \sigma_i \lambda_{SO})}^2}  \nonumber \\ \rm{and}\quad
 \tau_i &=& E_i  - (e l E_z - \eta \sigma_i \lambda_{SO})~,
\label{momenta}
\eea
with $\sigma_1=\sigma$, $\sigma_3=\sigma^{\prime}$, $E_1 = E+s h$, $E_3=E+s' h$ and $S_i = sgn [E_i-(e l E_z-\eta \sigma_{i} \lambda_{SO})]$.
Since momentum is conserved in the $y$ direction and does not change, it is convenient to write the $x$-component of the wave-vectors as 
\bea
  k_{xi}= \sqrt{{k_i}^2 - {k_{y}}^2} ~.
\label{momenta2}
 \eea
 where $k_{y}$ is the conserved momentum in the $y$ direction. 

In Eqs.~(\ref{momenta} and \ref{momenta2}) we consider $i=1,3$ only. For simplicity we assume that $E_z=0$ in the middle region which makes  the
momentum  in the middle region 
independent of valley and spin ($\sigma_{2}$). 
Hence for the central region, 
\bea
k_2 &=& \sqrt{{E_{2}^2 - \lambda_{SO}^2}}~~, \nonumber \\
k_{x2} &=& \sqrt{k_{2}^2 -{k_{y}}^2}~~, \nonumber \\
\rm{and}\quad \tau_{2} &=& E_{2} \ .
\label{momenta3}
\eea
where $E_{2} = E-U$ and $U$ is the height of the potential barrier  in the normal silicene region.

\section{Numerical results}  {\label{sec:IV}}
In this section we present our numerical results for the FNF junction for different parameter regimes. 
%
\begin{figure}
\centering
\includegraphics[width=1.0\linewidth]{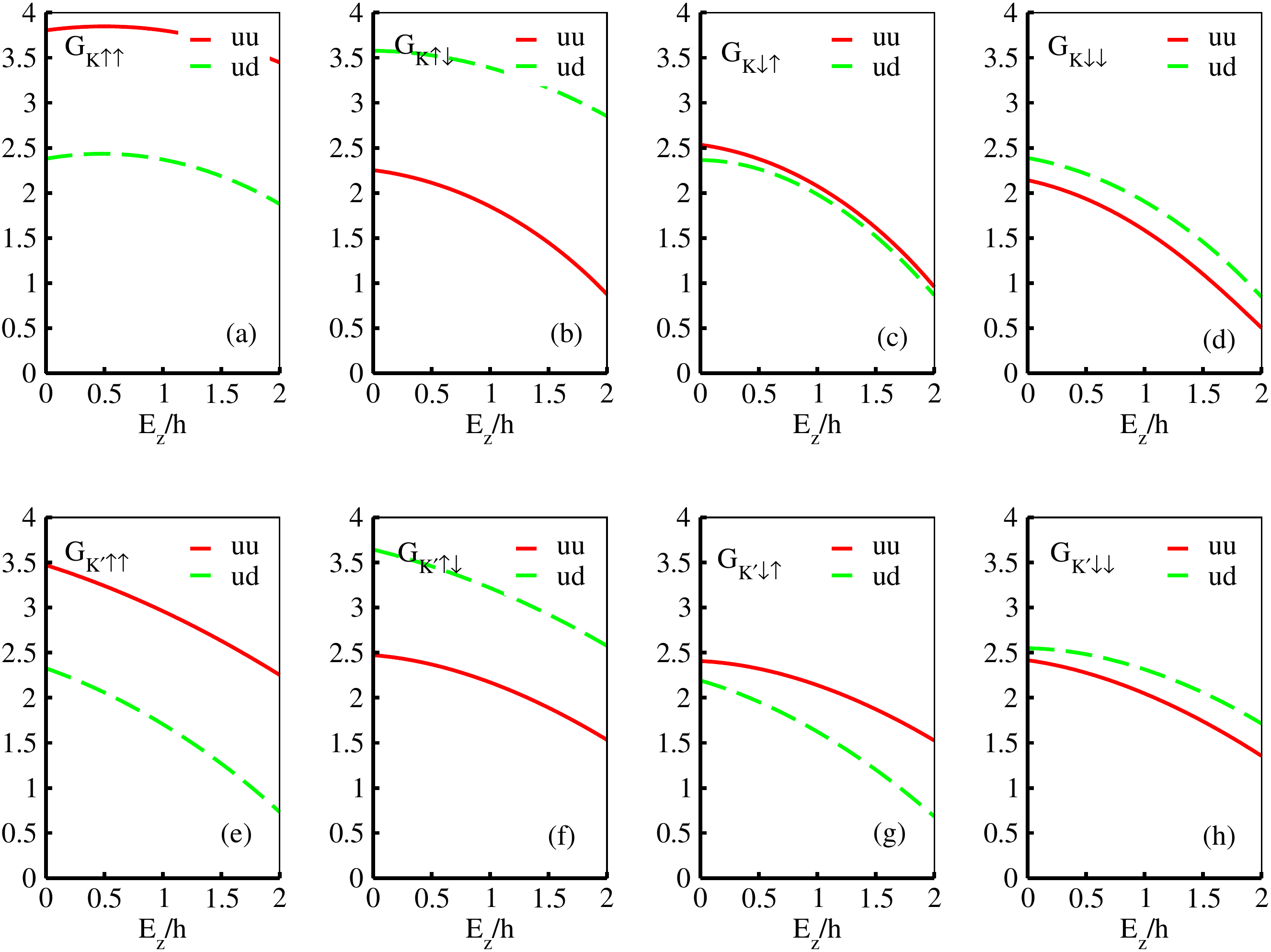}
\includegraphics[width=1.0\linewidth]{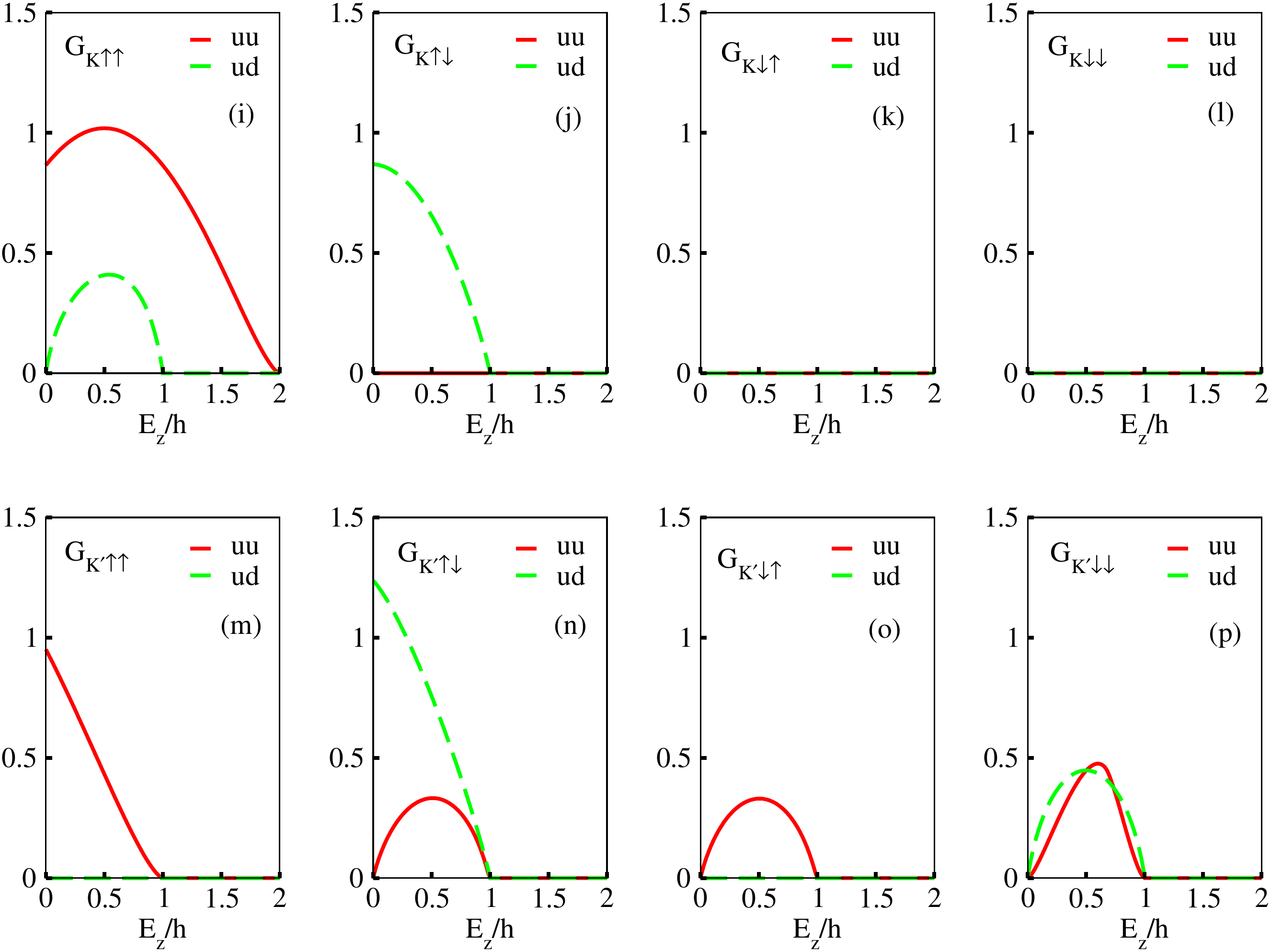}
\caption{(Color online) Conductances ($G_{V\sigma\sigma\rq{}}$) in units of $e^2W/\pi h$, for the P (uu)  and AP (ud)  configurations of a 
FNF junction  are shown as a function of the dimensionless parameter $E_{z}/h$ for $E>h$ [upper panels, (a-h)] and 
$E<h$ [lower panels, (i-p)] respectively. Here $E_{z}$ is the external electric field and $h$ is the ferromagnetic exchange field. 
The value of the other parameters are chosen to be $\lambda_{SO}/h=0.5 $, $U/h=30$. Energy of incident electron, for $E>h$ is
$E/h =4.0$ and for $E<h$ is $E/h=0.5$.}
\label{Fig4\rq{}}
\end{figure}
%

%
\begin{figure}
\centering
\includegraphics[width=1.0\linewidth]{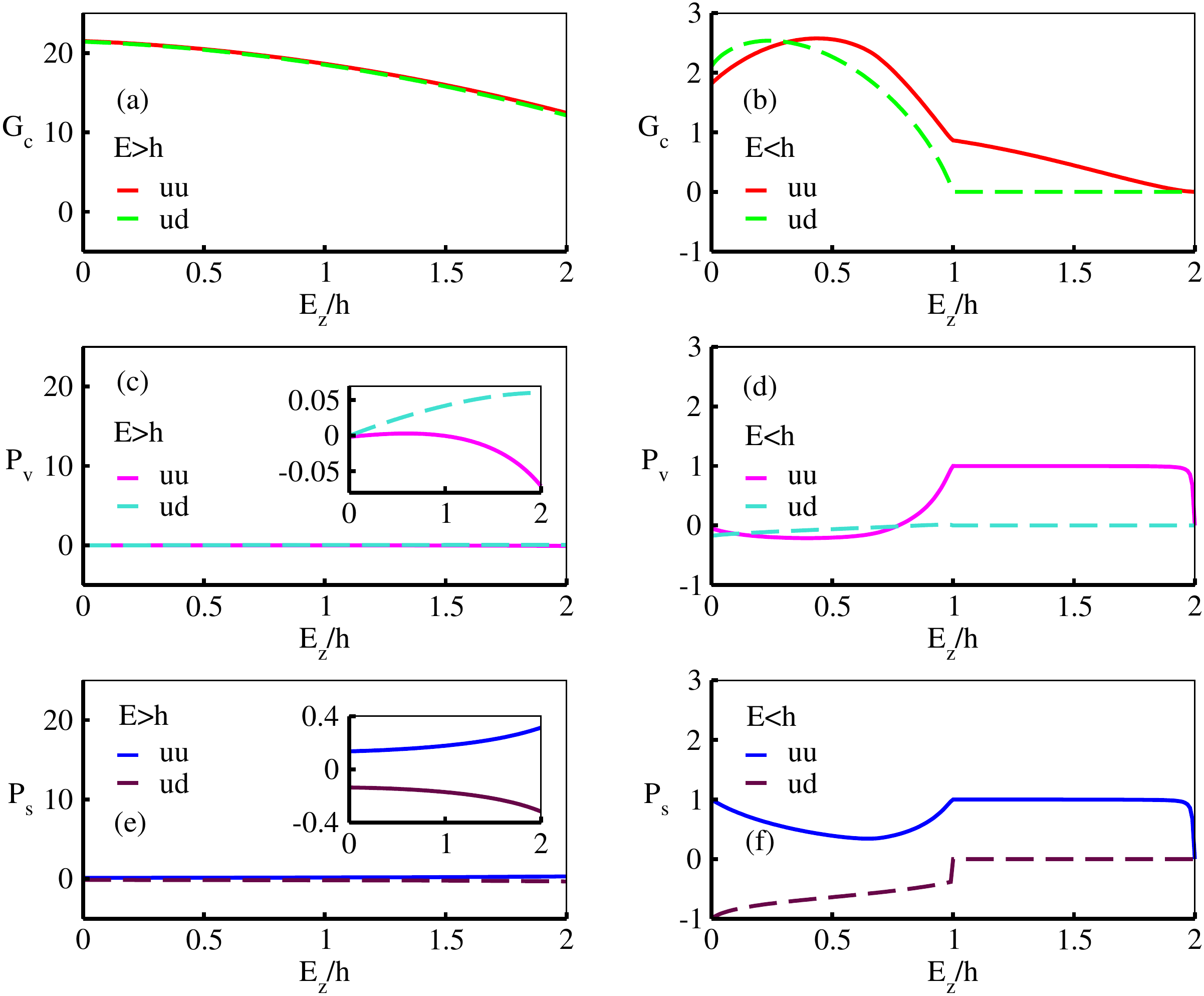}
\caption{(Color online) Total charge conductance ($G_{c}$) in units of $e^2W/\pi h$, valley polarization ($\mathcal{P}_{v}$) and spin polarization ($\mathcal{P}_S$) 
for P and AP configurations of a FNF junction are shown as a function of the dimensionless parameter $E_{z}/h$ for $E>h$ 
[left panels, (a), (c) and (e)] and $E<h$ [right panels, (b),(d) and (f)] respectively. The value of the other parameters are chosen 
to be the same as in Fig.~\ref{Fig4\rq{}}. The insets emphasize that $\mathcal{P}_{v}$ and $\mathcal{P}_S$ are actually different in 
magnitude for the $uu$ and $ud$ configurations for $E>h$ regime.}
\label{Fig4}
\end{figure}
%
We first study the model described in Eq.(\ref{hamiltonian}) which has a spin-independent barrier in the normal silicene region.
We compute the conductance using the transmission coefficients obtained in Eq.(\ref{matrix}) 
for both the parallel ($s = s\rq{}$) 
and the anti-parallel ($s = -s\rq{}$) 
configurations of spins, using the Landauer$-$Buttiker formalism~\cite{SDatta}. We use the scattering matrix to compute the total transmission 
probability $T^{ss\rq{}}(\theta_1)  = |t|^2 k_{x3}/k_{x1}$ for parallel and anti-parallel configurations by choosing the spins appropriately 
and for a particular incident angle $\theta_1$ (which fixes the angles in the other regions  as well). The factor of $k_{x3}/k_{x1}$ in the 
transmission function is needed because the probablity flux density includes a factor of the velocity which is essentially $\hbar k/m$. 
Since experimentally, it is not easy to control the angle of incidence of the impinging electron, 
we then compute the conductance by integrating over the possible angles of incidence and multiplying by the number of modes
within the width $W$ of the silicene sample.
At zero temperature, this leads us to a conductance given by 
\bea
G^{ss\rq{}}  = \frac{ {e}^2}{h} \frac{W k_{1}}{\pi} \int^{\theta_C}_0 T^{ss\rq{}}(\theta_1)\cos \theta_1 d \theta_1~.
\label{G}
\eea
Here, $\theta_C$ is the critical angle of the incident particles which is needed to ensure propagating particles
in the first and third regions and is given by $\theta_C = \pi/2 $ for $k_1 \le k_3 $
and
$\theta_C = \text{arcsin} (k_1/k_3) $ for $k_1> k_3$. 
Note that unlike the case for graphene, in silicene, the contributions at the two valleys are not identical and hence, we do
not get the degeneracy factor of two. Instead, the contributions at both the valleys have to be computed independently 
and added to obtain the total conductance through the junction.
Thus we define the total charge conductances $G_c$, valley and the spin polarizations ($\mathcal{P}_v, \mathcal{P}_S$) and TMR 
through the FNF junction in terms of the following constituent conductances$-$
$G^{ss\rq{}}_{V\sigma\sigma\rq{}}$. Here, $ss\rq{}$ denotes $uu$ (P) or $ud$ (AP) spin configurations, $V$ denotes the valley 
(${\bf {K}}$ or ${\bf{K}}\rq{}$)
and $\sigma$ denotes the spin of the incoming charge carrier in region 1 and $\sigma\rq{}$ denotes the spin of the outgoing
charge carrier in region 3, which can be different, because we have spin-orbit coupling in the system.
These conductances have been shown in Fig.~\ref{Fig4\rq{}} for both $E>h$ and $E<h$. It is now easy to realize,
in reference to the band diagrams given in Fig.~\ref{Fig3} that the conductances go to zero when there is a gap
in the density of states either in region 1 or 3. The maxima can also be understood by noting that the density of states
at those values of $E_z/h$ are maximum and reduce both when $E_z/h$ is reduced or increased. This can be checked for
each of the various conductances on a case by case basis.

The total charge conductance $G^{ss\rq{}}_{c}$ and the valley $\mathcal{P}_v^{ss\rq{}}$ and spin $\mathcal{P}_S^{ss\rq{}}$ polarizations 
for both the P ($s=s\rq{}$) and AP ($s=-s\rq{}$) configurations are now defined as 
\bea
G_{c}^{ss\rq{}} &=& \sum_{V\sigma\sigma\rq{}} G_{V\sigma \sigma^{\prime}}^{ss\rq{}}~,  \nonumber \\
\mathcal{P}_v^{ss\rq{}} &=& \frac {\sum_{\sigma\sigma\rq{}} (G_{\bf {K} \sigma \sigma^{\prime}}^{ss\rq{}}  - G_{\bf {K^{\prime}}\sigma \sigma^{\prime}}^{ss\rq{}}) } {G_{c}^{ss\rq{}}} ~,\nonumber \\
{\rm and} \quad  \mathcal{P}_S^{ss\rq{}} &=&\frac{ \sum_{V\sigma} (G_{V\sigma\uparrow}^{ss\rq{}} - G_{V\sigma\downarrow}^{ss\rq{}})} {G_{c}^{ss\rq{}}}~.
\label{GT}
\eea
Note that the indices $ss\rq{}$ in Eq.(\ref{GT}) gives rise to four possible spin configurations
$uu$, $ud$, $du$, $dd$ for a FNF junction, of which $uu$ and $dd$ imply P configurations with $s=s\rq{}$ and
$ud$ and $du$ denote AP configurations with $s=-s\rq{}$.

Now from the knowledge of all the possible conductances, one can define the tunneling magnetoresistance (TMR) 
through the FNF geometry as
\bea
\rm TMR &=& \frac{G_{c}^{uu} - G_{c}^{ud}}{G_{c}^{uu}} \ .
\label{TMR}
\eea
Note that the standard definition of TMR has $G_c^{ud}$ in the denominator. However it
is also sometimes defined with $G_c^{uu}$~\cite{diffdefinitionTMR} and we choose this definition because in our case, $G_c^{ud}$ vanishes at $E_z = h$. 
This implies a singularity in the TMR which is avoided in our definition. Note that for $E_z>h$, the difference in TMR between the 
two definitions is negligible.
For $E_z<h$, there are numerical differences, but no qualitative difference in the behavior of TMR with the two definitions.

The results are different for the energy regimes $E>h$ and $E<h$, because of the difference in band
structure, which has band gaps and hence no propagating states available for transport (see Fig.~\ref{Fig3}), 
for certain ranges of $E_z/h$ for $E<h$.
Since the main difference 
of silicene from graphene is the fact that the gap in silicene is tunable by the external electric field $E_z$,  we choose to 
focus on the dependence of conductances on  $E_z$.  In cases, where we study the conductances as functions of other
parameters such as the barrier strength $U$ or the exchange splitting $\Delta$, we present our results for three 
different values of $E_z/h$.

Note that when $E_{z}=\lambda_{SO}$, silicene is actually coplanar, \ie~the two sublattices are in same plane like in  graphene. 
But the spin-orbit coupling in silicene is much stronger than in graphene~\cite{CCLiu1}.  This increases or 
decreases the momentum of the incident charge carrier (see Eq.(\ref{momenta})) depending on the 
spin-polarization of the ferromagnetic silicene. Hence, we do not
expect to reproduce the results  of FNF junctions in graphene in the gapless regime. 
%
\begin{figure}
\centering
\includegraphics[width=1.0\linewidth]{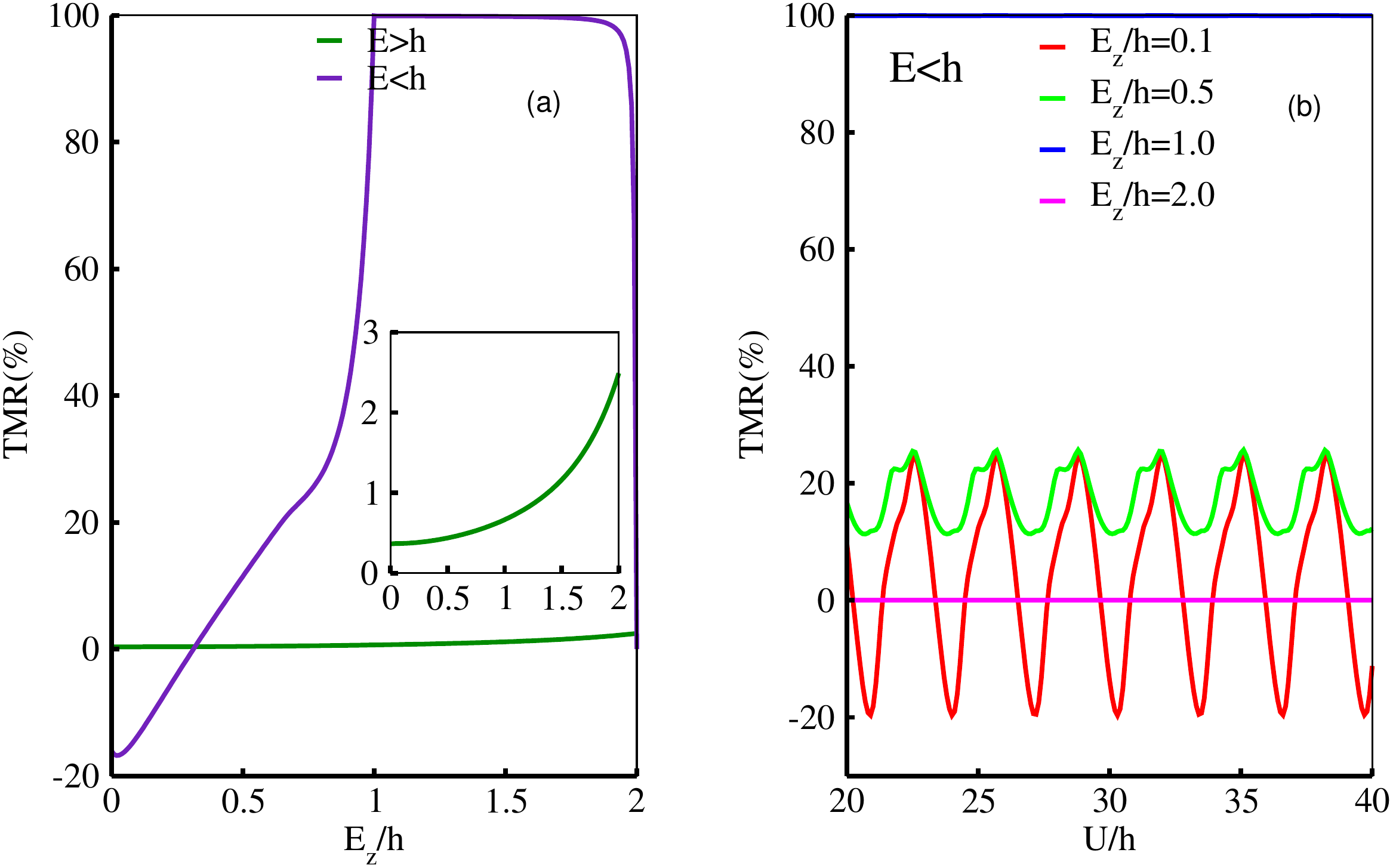}
\caption{(Color online) TMR is shown as a function of the dimensionless scale $E_{z}/h$ in panel (a) with
green and purple lines corresponding to $E>h$ and $E<h$ regime respectively. In panel (b) 
TMR is shown as a function of $U/h$ for $E<h$ for three values of $E_z/h$.
The other parameters are chosen to be the same as in Fig.~\ref{Fig4\rq{}}. In the inset of panel (a) we emphasize the very slow 
rate of increase of TMR in the $E<h$ regime.
}
\label{Fig5}
\end{figure}
%

\subsection{Spin-independent Barrier}
Here we discuss the case where we have a finite  spin-independent (scalar) barrier in the normal silicene region. 
The energy of the incident electron can be in the regime $E>h$ or $E<h$ and we present  the behavior
of the conductances and the TMR below. To carry out our numerical analyses,  we have chosen to normalise all our energy scales by the 
Zeeman energy $h$, so that all our results are in terms of dimensionless quantities. We also choose to measure conductances
in units of $e^2W/\pi h$.

We first present the results for the various constituent conductances for the P and AP configuration  in Fig.~\ref{Fig4\rq{}} in order to understand
the behaviour of the conductances, the valley and spin polarizations and the TMR in various parameter regimes. The conductances are shown
independently at the ${\bf K}$ and the ${\bf K}^\prime$ valleys as well as independently for the incoming ($\sigma$) and the outgoing ($\sigma'$) 
spins of the charge carriers.
In Figs.~\ref{Fig4\rq{}}(a-h), we show the behaviour of the conductances at the ${\bf K}$ and ${\bf K}^\prime$ valleys,
with respect to the dimensionless parameter $E_{z}/h$ for the four possibilities 
($\uparrow\uparrow, \uparrow\downarrow, \downarrow\uparrow, \downarrow\downarrow$) in the $E>h$ regime for
both the $uu$ and $ud$ configurations. Here the $uu$ 
configuration corresponds to the majority spin density of states in the left and right FS regions being up spin (parallel to each other)
and the $ud$ configuration corresponds to the majority spin being opposite (anti-parallel) in the two regions.
When $\uparrow$ charge carrier comes in from the left then it can either go to $\uparrow$ state or $\downarrow$ state in right region. 
So $\uparrow\uparrow$ etc, denote spins of the incoming and scattered charge carriers.
The behaviour of the various conductances in Figs.~\ref{Fig4\rq{}}(a-h) can now
be understood easily when analysed in terms of the band structure presented in Fig.~\ref{Fig3}. 

For $E>h$, the results have been
presented for $E/h=4$. Using the band diagram, it is easy to check that both at ${\bf K}$ and ${\bf K}^\prime$, there
are always electron states available for conductance for both the P and AP configurations and for all possible incoming and outgoing spins.
The differences in the magnitude
both at ${\bf K}$ and ${\bf K}^\prime$ valleys  stems from the decrease in the momentum of the propagating states at the Fermi energy, as can be
seen from the band diagram. 

The results for $E<h$ is shown in Figs.~~\ref{Fig4\rq{}}(i-p) for $E/h=0.5$.  
Consider the $uu$ case for the valleys ${\bf {K}}$ and ${\bf{K\rq{}}}$.
For $G_{{\bf K},{\bf K\rq{}}\uparrow\uparrow}$, (red lines shown in Fig.~\ref{Fig4\rq{}}(i) and Fig.~\ref{Fig4\rq{}}(m)),  
the band diagram (see Fig. ~\ref{Fig3}) shows that if we start with a spin up electron at
the $\bf K$ valley, (Fig.~\ref{Fig4\rq{}}(i)) then there is a non-zero density of states for $\uparrow$ electrons in the third region for all values of $E_z/h$ until
it reaches the value of 2 (the magenta line goes above the value of $E/h=0.5$). On the other hand, for the ${\bf K}^\prime$ valley, (shown
in Fig.~~\ref{Fig4\rq{}}(m)), there is no density of states for the $\uparrow$  electrons beyond $E_z/h=1.0$ (the blue line goes above the 
value $E/h=0.5$) in the third region. This explains why beyond $E_z/h = 2$ for the ${\bf K}$ valley and beyond $E_z/h=1$ for  the ${\bf K}^\prime$ valley, the conductances
$G_{{\bf K}\uparrow\uparrow}^{uu}$ and  $G_{{{\bf K}^\prime}\uparrow\uparrow}^{uu}$ are zero. It is also clear from the band diagram
that for $G_{{\bf K}\uparrow\uparrow}^{uu}$, its value increases from the value at $E_z/h = 0$,
because the momenta of the electrons at $E/h=0.5$ grows (comparing the red and green lines) and beyond that it decreases 
(comparing the green, blue and magenta lines). On the other hand, for $ G_{{{\bf K}^\prime}\uparrow\uparrow}^{uu}$, it is clear that the momentum
of the electrons at $E/h=0.5$  decreases as a function of $E_z/h$ (comparing the red, green and blue lines). This explains
why the conductance rises initially and then falls beyond $E_z/h =0.5$ for $G_{{\bf K}}^{\uparrow\uparrow}$ and 
why it falls monotonically for $G_{{\bf K}^\prime}^{\uparrow\uparrow}$. 
 
A similar detailed analysis can also be made for the $ud$ case as well as 
each of the other graphs in Figs.~\ref{Fig4\rq{}}(j,k,l) and Figs.~\ref{Fig4\rq{}}(n,o,p), which explains each 
feature of the graph. However, since the method is similar to what has been described above, we will not go through
each one of the graphs in detail. The behaviour of the  charge conductance, the valley and spin polarizations and the TMR 
are also now understandable, since
we can explain  how each of the constituents $G_{{V{\sigma\sigma'}}}^{ss'}$ behave as a function of $E_z/h$ from the band diagram.

%
\begin{figure}[h]
\centering
\includegraphics[width=1.0\linewidth]{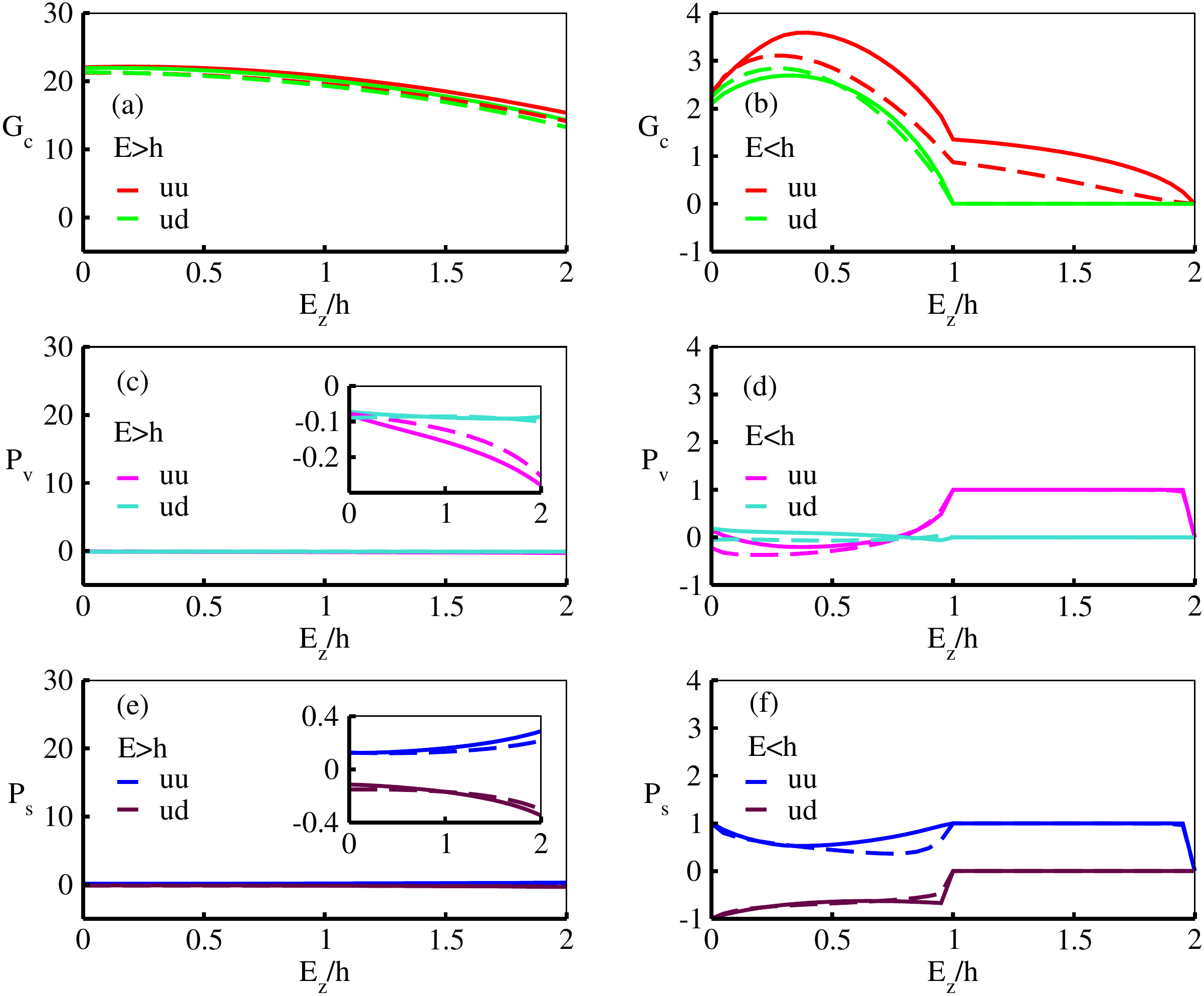}
\caption{(Color online) In the left panels, we show the total charge conductance ($G_{c}$) in units of $e^2W/\pi h$, valley and spin polarizations 
($\mathcal{P}_v^{ss\rq{}}$, $\mathcal{P}_S^{ss\rq{}}$) for $E>h$ and on the right panels for $E<h$ respectively. The dashed and solid lines
correspond to two different exchange splittings ($\Delta/h$) due to the spin-dependent barrier in the normal silicene region. 
The value of the other parameters are chosen to be $\lambda_{SO}/h= 0.5$, $U/h=30$. 
Energy of the incident electron, for $E>h$, is $E/h=4.0$ and for $E<h$, $E/h=0.5$.
The small difference in magnitude of $\mathcal{P}_v^{ss\rq{}}$ and $\mathcal{P}_S^{ss\rq{}}$ is highlighted in the insets for $E>h$ regime.}
\label{Fig6}
\end{figure}
%

\vspace{0.2cm}

\noindent \textsf{(a)}~$E>h$ 

In Figs.~\ref{Fig4}(a), \ref{Fig4}(c) and \ref{Fig4}(e), we show the behaviour of the charge conductance ($G_{c}^{ss\rq{}}$) and the valley 
and spin polarizations ($\mathcal{P}_v^{ss\rq{}}$
and $\mathcal{P}_S^{ss\rq{}}$)
with respect to the dimensionless parameter $E_{z}/h$ for the P and AP configurations 
($uu$ and $ud$) in the $E>h$ regime. Note that 
in Fig.~\ref{Fig4}(a), $G_{c}^{uu}$ and  $G_{c}^{ud}$
are  both finite at $E_{z}/h=0$ and start decreasing as we increase the value of $E_{z}/h$. $G_c$ is obtained by
summing $G_{\bf{K}}^{ss\rq{}}$ and $G_{\bf{K\rq{}}}^{ss\rq{}}$, which in turn are obtained as
\bea
 G_{\bf{K}}^{ss\rq{}} &=& (G_{{\bf{K}}\uparrow\uparrow}^{ss\rq{}}+ G_{{\bf{K}}\uparrow\downarrow}^{ss\rq{}}) +  
 (G_{{\bf{K}}\downarrow\uparrow}^{ss\rq{}} + G_{{\bf{K}}\downarrow\downarrow}^{ss\rq{}}) \nonumber \\
 &\equiv&  \sum_{\sigma\rq{}} G_{{\bf{K}}\uparrow \sigma^{\prime}}^{ss\rq{}} +G_{{\bf{K}}\downarrow \sigma^{\prime}}^{ss\rq{}}
\eea
and similarly for $G_{{\bf{K}}\rq{}}^{ss\rq{}}$. In other words, the total charge conductance $G_c^{ss\rq{}}=G_{\bf{K}}^{ss\rq{}}
+G_{{\bf{K}}\rq{}}^{ss\rq{}}$  is obtained by summing over all the conductances in the panels (a) to (h) in Fig.~\ref{Fig4\rq{}}. 

In Fig.~\ref{Fig4}(c), $\mathcal{P}_{v}^{uu}$ and $\mathcal{P}_{v}^{ud}$ are plotted which are close to zero
on the scale of the charge conductance. However, they are not identical, as shown in the inset. But
it appears that in this regime, silicene  has negligible  valley
polarization, similar to graphene, which in fact has no valley polarization at all, since the two valleys are identical. 
This can be understood because the valley polarization $\mathcal{P}_v$ is simply proportional to $G_{\bf{K}}^{ss\rq{}}-G_{{\bf{K}}\rq{}}^{ss\rq{}}$,
and as can be seen from Fig.\ref{Fig4\rq{}} that the magnitudes of the conductances at ${\bf {K}}$ and ${\bf{K}}\rq{}$ are almost the same for $E>h$.
In Fig.~\ref{Fig4}(e), the behaviour of the spin polarization has
been shown for both P ($uu$) and AP ($ud$) configurations, which is also very small in this regime. 
As shown in the inset, the spin polarization is positive for the P and
negative for the AP configurations and increases as a function of $E_z/h$.  This difference is due to the spin-orbit coupling in silicene, whereas
in graphene, they are much smaller, since the spin-orbit coupling is vanishingly small.
Finally, in Fig.~\ref{Fig5}(a), the behavior of the TMR is shown as a function of $E_{z}/h$ by the green solid line in the $E>h$ regime. 
In this regime, the TMR is close to zero.

\vspace{0.2cm}

\noindent \textsf{(b)}~$E<h$

The right panels in  Fig.~\ref{Fig4}  shows the  behaviour of the 
charge conductance $G_{c}^{ss\rq{}}$ and the valley and spin polarizations $\mathcal{P}_{v}^{ss\rq{}}$, 
$\mathcal{P}_{S}^{ss\rq{}}$ with respect to the dimensionless parameter $E_{z}/h$, in the $E<h$ regime, 
for the same spin and polarization configurations mentioned earlier. 
These are just the appropriate sums and differences of the
constituent conductances in Figs.~\ref{Fig4\rq{}}(i-p). Here also,
their behaviour is easy to understand by comparing each of the graphs in Figs.~\ref{Fig4\rq{}}(i-p) with the 
band diagrams in Fig.~\ref{Fig3} and noting when there is no density of states for the configuration in either the incoming 
or the outgoing spin configuration of the charge carriers. For instance, for the $uu$ case, there is only one contribution for $E_z/h>1$ 
in Fig. ~\ref{Fig4\rq{}}(i) and for 
the $ud$ case, there is no contribution for $E_z/h>1$.  This is because we have chosen $E/h=0.5$ and the blue line ($E_z/h=1.0$)
in the band diagram goes above that line either for the incoming or scattered region for all cases in the $ud$ configuration
and all but one case in the $uu$ configuration.
In other words, their behaviour follows what is expected from the 
availabality or non-availability of propagating states at the $\bf K$ and ${\bf K}^\prime$ valleys as explained above in the discussion
of Fig.~\ref{Fig4\rq{}}.

The most interesting point to note is that the valley polarization  and the spin polarization 
for the parallel or $uu$ configuration is unity for $E_z/h>1$ in $E<h$ regime. This is simply because of the entire
contribution to the conductance in this regime originates from $G_{{\bf{K}}\uparrow\uparrow}^{uu}$. So the
conductance is both fully valley and spin polarised and would be an important regime to achieve by tuning
the incident electron energy $E/h <1$ and the electric field $E_z/h>1$.
In the anti-parallel or $ud$ regime, the spin polarization can be tuned to negative values
when $E_z/h <1$, but without any valley polarization.

\vspace{0.5cm}
%
\begin{figure}[h]
\centering
\includegraphics[width=1.0\linewidth]{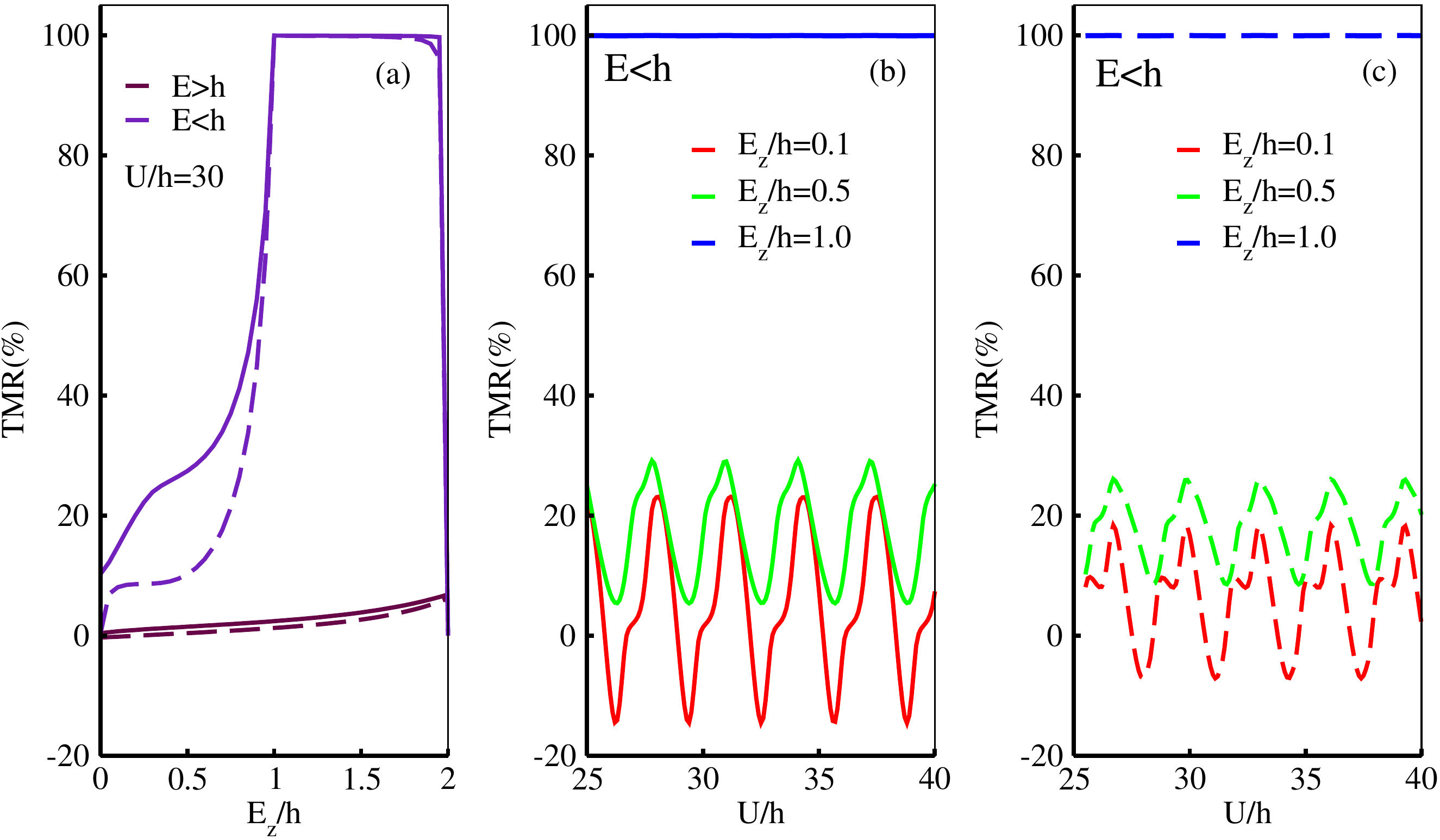}
\caption{(Color online) The variation of the TMR with respect to $E_{z}/h$ and $U/h$ is shown for  two values of $\Delta/h$. 
In the panels, the dashed and the solid lines correspond to $\Delta/h=0.5$ and $\Delta/h=-0.5$ respectively.
We choose the same value of the other parameters as mentioned in Fig.~\ref{Fig6}.
}
\label{Fig7}
\end{figure}
%

The behaviour of TMR is demonstrated in Figs.~\ref{Fig5}(a) and \ref{Fig5}(b) for the $E<h$ regime. For $E_{z}/h=0$, the TMR is negative
and reaches its maximum negative value. 
Then the TMR increases as we increase the value of $E_{z}/h$ and changes sign and reaches saturation at $E_z/h \simeq 1$ since the conductance
becomes fully spin polarised at that point. Note that 
we need  to restrict the value of $E_{z}/h$ below two as the TMR takes an indeterminate form at $E_{z}/h=2$ 
due to the vanishing of both 
$G_{c}^{uu}$ and $G_{c}^{ud}$ (see Fig.\ref{Fig4}(c) and Eq.(\ref{TMR})) for all spin configurations. This is a consequence of our  
choice of the incident energy at $E/h=0.5$.  

The striking feature of positive to negative transition in the TMR also arises as we vary the strength of
the potential barrier $U$ in the middle normal silicene region for different values of $E_z/h$.
This feature is shown in Fig.~\ref{Fig5}(b) where the TMR 
oscillates between positive and negative values with respect to $U$ for $E_{z}/h=0.1$.  
Such oscillations of the TMR from positive to negative values have 
been reported earlier in Ref.~\onlinecite{JZou} in 
graphene, due to the change in the type of the charge carrier in the third region.
Note also there is no significant qualitative change in the 
behavior of TMR even if we choose $U\sim h$. 
This  extra tunability of the TMR 
with respect to an external electric field is a unique feature of silicene that we wish to emphasize here in this manuscript.

\subsection{Spin dependent Barrier}
Here we discuss the effect of a spin-dependent barrier on the total charge conductance, valley and spin polarizations and the TMR. 
The barrier is modelled in the normal silicene region as $U_{\sigma} = U - \sigma \Delta$ which is shown in Fig.~\ref{Fig2}. 
Here positive (negative) $\Delta$ represents the exchange splitting in the silicene barrier with its magnetization parallel (anti parallel) 
to the spin orientation of the FS in the first region.

In Figs.~\ref{Fig6}(a-f) we show the behaviour of the total charge conductances $G_{c}^{ss\rq{}}$, spin and valley polarizations 
$\mathcal{P}_{S}^{ss\rq{}}$ and $\mathcal{P}_{v}^{ss\rq{}}$, in the $E>h$ and $E<h$ regimes, 
for $\pm \Delta/h$. Since the qualitative behavior of all the conductances remain similar to the spin independent barrier case,
we do not show the behaviour of the conductances at the ${\bf K}$ and ${\bf K}^\prime$ valleys independently, or analyse the
graphs in detail via the band structure.  Similarly, in Figs~\ref{Fig7}(a-c), we show the behaviour of the TMR as a function of
$E_z/h$ and as a function of $U/h$ as well (for $E_z<h$), for both $\pm \Delta/h$. We find that the results are fairly
similar to the spin-independent barrier case.

\section{Summary and conclusions} {\label{sec:V}}
To summarize, in this paper, we have investigated the transport properties (charge conductance as well as spin and valley polarizations) 
and the TMR through a FNF junction in silicene. Here we have adopted the Landauer-Buttiker formalism to carry out our analysis. 
We show that the conductances and the TMR 
in this geometry can be tuned by an external electric field $E_{z}$ for each case
($\uparrow \uparrow$, $\uparrow\downarrow$, $\downarrow\uparrow$, $\downarrow\downarrow$) in the left and right ferromagnetic silicene regions,
for both parallel ($uu$ or $dd$) and anti-parallel ($ud$ or $du$) configurations. 
For specific values of the electric field, we analyse both the charge conductance and valley and spin polarizations in terms of the independent
behaviour of the conductances at the two valleys and the band structure at specific incident energies.
We find that we can tune a fully valley polarised and also a fully spin polarised current through our setup via the external electric field. 
We also find that the TMR can be tuned to 100 \% in this geometry via the electric field.
This is one of the main conclusions of our analysis.
We also show that the TMR through our setup exhibits an oscillatory behavior as a function of the strength of the barrier 
(both spin independent and spin dependent) in the normal silicene region. The TMR also changes sign between positive and  
negative values and such a transition can be tuned by the external electric field. This is another conclusion of our analysis.
Hence, from the application point of view, our FNF geometry may be 
a possible candidate for making future generation spintronic devices out of silicene.  

As far as the practical realization of such a FNF structure in silicene is concerned, it should be possible to
fabricate such a geometry with the currently available experimental techniques.
Ferromagnetic exchange in silicene may be achieved via proximity effect using a magnetic insulator, for instance, EuO~\cite{hxyang,Brataas}.
The typical spin orbit energy in silicene is $\lambda_{SO}\sim 4~\rm meV$~\cite{CCLiu1}. For an incident electron with energy 
$E\sim 4~\rm meV$ and exchange energy $h\sim 8~\rm meV$, the maximum value of the spin and valley polarization as well as 
sign change of TMR from positive to negative value occur at an electric field $E_{z}\sim 0.03~\rm V \mathring{A}^{-1}$, potential barrier 
of height $U \sim 160~\rm meV$, exchange splitting in the normal silicene region $\Delta\sim 4~\rm meV$ and width of the barrier 
$L\sim 100~\rm nm$.

\section*{Acknowledgments} 
One of us (A.S.) would also like to acknowledge the warm hospitality at the University of Basel, Switzerland,
where this work was initiated.

\vspace{2cm}

\appendix

\bibliography{Silicene_FNF_ref} 

\end{document}